# Are Temperature and other Thermodynamics Variables efficient Concepts for describing Granular Gases and/or Flows ?

## P. Evesque

Lab MSSMat, UMR 8579 CNRS, Ecole Centrale Paris
92295 CHATENAY-MALABRY, France, e-mail: evesque@mssmat.ecp.fr

**Abstract:**

*Granular flows and vibro-fluidised granular gases have been extensively studied recently; most of the theoretical analyses and the experimental descriptions use temperature and other thermodynamics concepts. However, taking the very simple case of a vibro-fluidised gas made of identical particles, we show the lack of efficiency of such concepts for the understanding of the physics of such systems. This results from both (i) the fact that the vibrator does not transmit the same amount of energy to each particle, but an amount which depends on its mass and/or its size and (ii) from the fact that it is a strongly dissipative medium. We conclude that most experimental device works rather as velostat as a thermostat.*

**Pacs # :** 5.40 ; 45.70 ; 62.20 ; 83.70.Fn

______________________________________________________________________

## 1. Introduction

Indeed, many works are dealing with granular gases and granular flows nowadays (see for instance [1-5] and refs. in [2]); and most of them [1-3] are introducing granular temperature and other thermodynamics concepts to describe their results and to measure some physical quantities, so that these concepts are supposed to be adequate for describing the physics of these media.

On the other hand, it has been also commonly recognised by the same (and other) workers, that granular materials are far from being at thermodynamics equilibrium, especially when the quasi-static regime is involved, *i.e.* when these materials are dense and compact. In the same way, one knows that granular gases and flows are strongly dissipative so that they shall be also far from thermodynamics equilibrium. So, a question arises: Can one apply thermodynamics concepts to these media without any care?

This paper tries answering this question. It excludes the case of quasi-static because no real thermodynamics description has been tempted already; this does not mean it should not be tempted in this case, because the validity of this approach could just arise from the hyperstaticity of the material, which might generate a large degeneracy of states and make a statistical treatment efficient. Anyhow, this is not the problem treated here, for which we limit to dynamical problems. Furthermore, we restrict the problem to the limit of granular gases, since it seems at first sight that this peculiar case should be the archetype domain where thermodynamics concepts should apply very easily and where the classic kinetic theory of gas should work well.

So, in order to answer this question, we investigate the property of a dilute vibro-fluidised granular gas under gravity, submitted to strong vibration; as the particles move fast and randomly in this specific range, one can think that ergodic motion is ensured and that the granular gas is described efficiently by rules issued from thermodynamics and from kinetic theory of gas. However, this paper will demonstrate just the contrary: First of all we will show that a vibrator is not a good thermostat for granular gas. Secondly we will show that the mixing of two gases of different particles leads to some problem of segregation most likely; such a segregation shall arise already when the particles have the same size but different masses. Then we will investigate the physical properties of a gas of identical particles as a function of the size and of the mass of its particles; we will argue that the effect of particle size should no be so important in a given range of excitation, if the number of layers making the gas remains constant; this will allow to compare our theoretical approach to the nice experimental results of ref. [3] and to prove the validity of our approach..

Indeed, paper [3] seems to lead to puzzling physics, since it demonstrates that the dynamics of a granular medium made of two kinds of grains and submitted to a vertical vibration is characterised by two different temperatures, one for each kind of grain; and that these temperatures depend on the proportion of each particle. But we will show that it simply confirms the present approach which demonstrates that a vibrator operates rather as a **"velostat"** than as a **"thermostat"**, which means that it imposes a velocity and not a kinetic energy. Hence, it shows the contradiction with classic thermodynamics approach as soon as one tries to mix different particles.

This allows to conclude that thermodynamics quantities and concepts, such as temperatures,…, might be more difficult to transpose to the physics of granular gas and granular flow than it was thought originally [1, 2]; this difficulty comes likely from both (i) the difficulty of building thermostats and other thermodynamics tools, and (ii) from the local dissipation which makes the granular gas non homogeneous and granular gas quantities non conservative [3-5] so that these systems are quite far from equilibrium.

These general trends will be tentatively extended in the conclusion to other domains of dissipative systems; we will mention in particular the case of shear flows for which dissipation and transfer of energy are both strongly local and inhomogeneous, depending on the particle size and particle size distribution. But a completely different domain of application will be also briefly invoked; it is the domain of economics, because taxes and redistribution can be seen as the governing laws of dissipation occurring during transfer of goods; they work at the very local scale and may lead to inhomogeneous behaviour. May these results explain the heterogeneous structure observed in economics, with small and large company and explain the "fractal" nature of the distribution of economic participants, of economic transfer….

## 2. A mechanical vibrator operates as a "velostat":

Let us demonstrate that a vibrator acts as a **"velostat"** rather than as a **"thermostat"**. We consider first the case of a granular gas made of a single kind of particles of diameter d and mass m submitted to a vertical sinusoidal vibration (frequency f=$\omega/(2\pi)$, amplitude a) imposed by a piston (mass M, horizontal size L). Be also nd the mean thickness of the layer when a=0; so, n stands for the mean coverage ratio, *i.e.* n is the mean number of bead layers covering the piston at rest; and the total number of particles N is given by N=$\alpha$n(L/d)², where $\alpha$ is the coefficient of normalisation which takes into account the density of arrangement of the regular lattice used for normalisation.

Let us now assume :
   (i)    that L>>d and L>>d/$\sqrt{n}$, so that lateral wall effects become negligible ;
   (ii)   that M/L² >>m/d² and M/L² >>n m/d² so that the piston mass can be considered as infinite for collision rules;
   (iii)  that bead-bead and bead-piston collisions are both governed by the same restitution coefficient $\varepsilon$ ;
   (iv)   that we can neglect the air effect and its viscous drag;
   (v)    Let us also neglect particle rotation, stating for instance that the rotation kinetic energy of a grain behaves as its mean translation kinetic energy.

The equality between inertial mass and gravity mass imposes that the mass m of the particles does not play any part in determining the particle -trajectories and -speeds because all masses are equal so that mass intervenes neither in the free-flight dynamics nor in the collision characteristics, as far as the restitution coefficient $\varepsilon$ remains independent of the speed V. This leads to predict that two gases of identical particles having the same diameter and the same coverage ratio n and submitted to the same vibration excitation vibrate similarly even if the two particles have different masses $m_1$ and $m_2$, the restitution coefficient being assumed the same, *i.e.* $\varepsilon_1=\varepsilon_2$. It means that the particles of these two gases get the same typical speed V , independently of m. So, V=$V_1$=$V_2$ . Of course, V depends on the other experimental parameters which are a, $\omega$, $\varepsilon$, n, g, d; in particular, one expects that the larger the amplitude a and/or the pulsation $\omega$ the larger the grain speed so that dimensional analysis leads writing:

$$V= a\omega \, F\{\varepsilon, n, d/a, gd/(a\omega)^2\} \qquad (1)$$

where F is a function of 4 dimensionless parameters. Indeed, as m/m=1 and m/M=0, they are both independent of m; and dimension analysis confirms that Eq. (1) is independent of m. Furthermore, as the "granular temperature" T is defined as T=$<mV^2/2>$= $mV^2/2$, this demonstrates that a vibrator is a **"velostat"** instead of a **"thermostat"**, which means that **it imposes a speed** and not a temperature (or a mean kinetic energy). Indeed if we consider two granular gases differing only by the mass $m_1$ & $m_2$ of their grains, so that $m_1\neq m_2$ , but $d_1=d_2$ , $n_1=n_2$ , $\varepsilon_1=\varepsilon_2$ , $a_1=a_2$ and $\omega_1=\omega_2$ , Eq. (1) imposes $V_1=V_2$ so that their temperatures $T_1$ & $T_2$ are different, their ratio being $T_1/T_2=m_1/m_2$. But the heights $H_1$ and $H_2$ of the two clouds are equal under the same external excitation, *i.e.* $H_1=H_2$, since $V_1=V_2$.

## *2.1. Mixing of two gases of different masses:*

Let us now investigate the mixing of these two gases in proportion $c_1$ & $c_2$, $c_1+c_2=1$, keeping $n=n_1 c_1+n_2 c_2$ constant: be $m_1>m_2$ and $T_{1,o}$ & $T_{2,o}$ their temperatures when they are separated; so $T_{1,o} > T_{2,o}$ but $V_{1,o} = V_{2,o}$; furthermore as one expects that mixing shall smoothen the temperature difference, one expects then that mixing (i) lets $T_1>T_2$, but that (ii) $T_1-T_2 < T_{1,o}-T_{2,o}$, so that mixing heats up the lighter particles from $T_{1,o}$ to $T_1=T_{1,o}+\Delta T_1$ and cools down the heavier ones from $T_{2,o}$ to $T_2=T_{2,o}-\Delta T_2$ in such a way (i) that $\Delta T_1 + \Delta T_2 < [T_{1,o}-T_{2,o}]$ and (ii) that the larger $c_2$ the larger $\Delta T_1$ and the smaller $\Delta T_2$, while the smaller $c_1$ the larger $\Delta T_2$ and the smaller $\Delta T_1$. In particular, this analysis does not conclude to the necessity of a thermodynamics equilibrium between the two species so that one shall accept that $T_1$ be different from $T_2$ in general. Furthermore the two mean speeds $V_1$ & $V_2$ are now different; so one expects that $H_1$ and $H_2$ be different too, with $H_1<H_2$, *leading probably to predict the existence of some spatio-temporal segregation*.

## *2.2. Comparison of two gases with different masses and diameters:*

Let us now compare the characteristics of two granular gases of identical particles under the same excitation $(a,\omega)$; as we want to investigate the gaseous case, we limit to small value of n and rather large value of the acceleration $a\omega^2>>g$. We consider two granular gases with the same number of layers at rest, so that $n_1=n_2$; furthermore $n_1=n_2 \approx 1$ because the medium looks like a gas. Be $(m_1,d_1)$ & $(m_2,d_2)$ the particle- (mass, size) of each gas respectively. Eq. (1) predicts that their speed $V_1$ and $V_2$ can be different, due to the difference between $d_1$ and $d_2$. However, let us assume that we are investigating the case of a granular gas for which the cloud height H shall be large, *i.e.* H>>d, H>>a. Indeed in this limit, one expects that confinement is due to g, so that H scales as $V^2/g$ and that V scales as $a\omega$, for a fixed set (n, d, m). This imposes that d/a and $gd/(a^2\omega^2)$ play only little part, and we will be neglect these dependence. In turn this assumes that $a\omega^2>>g$. In other words, we assume here that the typical speed of the gas depends essentially on the restitution coefficient $\varepsilon$ and on the number of layers n, but very little on the mass and the size of the grains. This is because the mass m does not intervene in the dynamics of identical particles and because the size d of the grain does not seem to have to play any part in the collision process, for a given number of layer, as far as the length L of the cell is large, *i.e.* L>>d, L>>H.

Indeed, it is worth noting by passing at this stage that the mean free path $l_c$ is given by $l_c=(L^2 H)/[N\pi d^2]$ so that it scales as H/n for which d does not play any part. Furthermore, when n>1, one expects the gas to be not homogeneous along the vertical so that $l_c$ is not constant with the coordinate z but increases with the height z in the cloud; this is due to both the action of local collision dissipation and to the mean confining pressure which decreases when the upper layer is approached.

## 2.3. Extension to the mixing of two gases with different mass and slightly different diameters : Comparison with experiment

According to the previous approximation, as far as n remains constant, one shall expect that mixing of two granular gases made of particles of different sizes shall behave approximately as the mixing of two granular gases of identical particles having different masses. So, one can compare this prediction to the experimental results of paper [3]; indeed, in this paper the ratio of the two particle masses $m_1/m_2$ is given by $m_1/m_2=(d_1/d_2)^3=(5/4)^3=1.95$, and the 3 proportions $(c_1,c_2)$ which was studied in this paper preserve the constancy of n, since the bead number $(N_1,N_2)$ satisfy $(N_1d_1^2+N_2d_2^2)= (c_1N_{1o}d_1^2+ c_2N_{2o}d_2^2)= (700\ d_1^2) = (1080\ d_2^2)= \alpha L^2$ = constant, with $N_{1o}=700$ and $N_{2o}=1080$. So taking into account the mass ratio $m_1/m_2=1.95$, the approach of §-2.1 predicts that $T_2$ shall decrease continuously from 1.95 to 1 when increasing $c_2$ from 0 to 1; it predicts also that $T_1$ shall decrease when $c_2$ increases and that $T_1/T_2<T_{1,0}/T_{2,o} =1.95$. Indeed, Fig. 3 of *ref.* [1] displays these behaviours: the larger the value of $c_2$ the cooler $T_1$ and the hotter $T_2$ (even if $T_2$ does not vary so much, *i.e.* $T_2=2.6 \cdot 10^{-5}$ J); this is in agreement with the previous approach. Furthermore as $T_2$ remains constant about, rather independent of $c_2$, one shall assume that $T_{2,o} \approx 25 \cdot 10^{-6}$J $\approx T_2$. This value of $T_{2,o}$ allows to predict $T_{1,o}=(1.95\ T_{2,o}) \approx 49 \cdot 10^{-6}$ J . This is just compatible with the observed data of Fig. 3.

So our approach seems to be in good agreement with the experimental results; it seems also to be in better agreement with these results than the values reported in Table 1 of ref. [3] ; so this cast a serious doubt on the validity of the theories which have allowed to establish this Table 1.

On the other hand, our approach is not able to predict the constancy of $T_2$ when $c_2$ increases. This demonstrates that much work has to be performed in this stimulating domain of the granular gases. Is this constancy linked to a $d_2/d_1$ dependence?

Let us now pursue and describe further consequences of this approach. The scaling which has been used shall be valid in the small g limit. So it shall be valid for g=0 too, which means for weightlessness condition. Making g=0 in Eq. (1), one gets that the typical speed V of a mono-disperse gas shall depend on n and ε only, if one neglects the effect of d. So, taking $T_{2,o} \approx 25\cdot10^{-6}$ J $\approx T_2$, and density $\rho=8\cdot10^3$ kg/m$^3$, one finds the typical speed $V_{2,o}=V_{1,o} = [3T_{2,o}/(16\pi\rho d^3)]^{1/2} = 0.216$m/s; this can be rewritten as $V_{2,o}=V_{1,}=(0.413\ a\omega)$, which fixes the value of the function F, *i.e.* F=0.413, with this set (n,ε) of parameters. This value for F, F=0.413, is not far from the value 0.25 which has been found approximately with the experiment on granular gas of bronze beads with n≈1 about in weightlessness condition [4, 5] in a vibrating cubic container. The difference can be explained by the difference of restitution coefficient and of n.

At this stage, it is also worth recalling one of the main result of ref. [5] because it may enlighten the discussion on the validity of thermodynamics concepts applied to granular gases: this main result is that a homogeneous granular gas has been only observed when the number of layers n is smaller than 1; when it becomes larger than 1 one observes the formation of a cluster: this means that the homogeneous granular gas exists only in the so-called Knudsen regime for which $l_c>L$, and for which no continuous mechanical description holds [6]. This tends to prove the failing of any simple continuous mechanical approach for granular gas. Does this result come from the fact that these media are strongly dissipating? Does it means also that a continuous formalism can only be based on a fractal description generating inhomogeneities at all scales of description? Does this mean also that the physics of a granular gas looks much more like the physics of a second order phase transition, for which large-scale fluctuations dominates and for which heterogeneity plays at all scales play a determining part? Future works will tell.

## 3. Conclusion: Dissipation and flow; dissipation and economics

The main conclusion of this work will be brought by the following remark: this simple approach demonstrates that granular gases violate the classic laws of the kinetic theory of gas: this is because it is not easy to build a good thermostat for a "granular medium" so that the temperature of a mixture is not a well defined quantity. A **"velostat"** seems to be an easier tool, which can be built at least when particles are identical; but the concept fails also at once when the granular gas is made of different particles of different masses. So the real question is: (1) Are classic thermodynamics concepts and variables efficient to describe granular gases? Or do these quantities so inhomogeneously distributed that they become meaningless? (2) How strongly does local dissipation perturb the problem and make the system inhomogeneous at all scales? Future works are indeed needed to answer this interesting questioning.

Can these objections apply also to other domains of granular physics; probably yes: for instance the "thermal agitation" of the grains in a flow is related to shearing; but shearing imposes a distribution of speed which depends of the gradient of deformation per unit time and on the particle size. So one expects that the fluctuations of speed (i) are local, (ii) do not depend on the grain mass at first approximation, (iii) but depends on the grain size so that flow-shearing is more a **"velostat"**, rather than a "thermostat". But the characteristic of this velostat is also local, linked to the local velocity and to the local grain size. As shearing generates also local dissipation, it can occur that main thermodynamics concept fail to be useful at describing such flows.

An other important point which makes the thermodynamics approach problematic in most applications on dissipative systems is the dissipation itself. Indeed, the effect of this dissipation is so efficient that it casts some doubt on the existence of a possible continuous field approach in the case of granular gas; this is because the gradients of physical quantities which are generated due to dissipation may vary as fast as the local density, so that the whole problem is self sustained. In order to exemplify this point we have recalled a recent result from our weightlessness experiment on vibro-fluidised granular gas: this experiment has shown that the gas is observed only when the mean free path $l$ of the particles is of the order of the sample size L, whereas a cluster is formed as soon as the particles collide. This cluster formation demonstrates the breaking of the homogeneous medium as soon as L>$l_c$. On the other hand, one knows from the framework of the kinetic theory of gas, that equations of continuous mechanics require that the representative elementary volume be larger than the mean free path $l$; this requires that the volume L of the samples shall be larger than $l$, *i.e.* L>$l_c$ , to derive a continuous field approach. So our weightlessness experiment seems to exhibit some incompatibility with the field approach.

This paper shows the difficulty to apply thermodynamics concepts to dissipative systems. It shows also that heterogeneous structure are generated in many cases. Of course this shall be a general result which might apply to other fields. A domain in which such fundamental concepts may apply could the domain of economics, for which all kind of taxes and of redistribution is some kind dissipation which inhibits transfers of goods, of information, of works. It might be that the strongly heterogeneous distribution of company size, of city sizes, … could come from this dissipation. In this case one should think a little more before tempting to apply the so-called Tobbin tax and other new concepts.

*Acknowledgements:* CNES is thanked for partial funding.

The electronic arXiv.org version of this paper has been settled during a stay at the Kavli Institute of Theoretical Physics of the University of California at Santa Barbara (KITP-UCSB), in june 2005, supported in part by the National Science Fundation under Grant n° PHY99-07949.

*Poudres & Grains* can be found at :
http://www.mssmat.ecp.fr/rubrique.php3?id_rubrique=402